\documentclass[aps,pra,twocolumn,superscriptaddress,amsmath,showpacs,amssymb]{revtex4} 

\usepackage{graphicx}
\usepackage{amsthm,amsmath}
\usepackage{color}

\newcommand{\bm}[1]{\mbox{\boldmath{$#1$}}}
\newcommand{\be}{\begin{eqnarray}}
\newcommand{\ee}{\end{eqnarray}}

\newcommand{\nn}{\nonumber}

\newcommand{\bo}{\boldsymbol}
\newcommand{\lb}{\label}

\newcounter{ichi}
\newcounter{ni}
\newcounter{san}
\newcounter{yon}
\newcounter{go}
\newcounter{roku}
\newcounter{nana}
\newcounter{hati}
\newcounter{kyu}
\begin{document}

\preprint{APS/123-QED}

\title{Quantum phases in $p$-orbital degenerated attractive 1D fermionic
optical lattices}

\author{Keita Kobayashi}
\affiliation{CCSE, Japan Atomic Energy Agency, 5-1-5 Kashiwanoha, Kashiwa, Chiba 277-8587, Japan}
\author{Yukihiro Ota}
\affiliation{CCSE, Japan Atomic Energy Agency, 5-1-5 Kashiwanoha, Kashiwa, Chiba 277-8587, Japan}
\author{Masahiko Okumura}
\affiliation{CCSE, Japan Atomic Energy Agency, 5-1-5 Kashiwanoha, Kashiwa, Chiba 277-8587, Japan}
\author{Susumu Yamada}
\affiliation{CCSE, Japan Atomic Energy Agency, 5-1-5 Kashiwanoha, Kashiwa, Chiba 277-8587, Japan}
\affiliation{Computational Materials Science Research Team, RIKEN AICS, Kobe, Hyogo 650-0047, Japan}
\author{Masahiko Machida}
\affiliation{CCSE, Japan Atomic Energy Agency, 5-1-5 Kashiwanoha, Kashiwa, Chiba 277-8587, Japan}
\affiliation{Computational Materials Science Research Team, RIKEN AICS, Kobe, Hyogo 650-0047, Japan}

\date{\today}% It is always \today, today,

\begin{abstract} 
We examine quantum phases emerged by double degeneracy of $p$-orbital
 bands in attractive atomic Fermi gases loaded on a 1D optical lattice. 
Our numerical simulations by the density-matrix renormalization group 
predict the emergence of a state with a charge excitation gap, the
 Haldane insulator phase. 
A mapping onto an effective spin-$1$ model reveals its physical origin. 
Moreover, we show that population imbalance leads to richer diversity of
 the quantum phases, including a phase-separated polarized state. 
Finally, we study the effects of harmonic trap potential in
 this 1D chain.
\end{abstract}

\maketitle

%%%%%%%%%%%%%%%%%%%%%%%%
\section{Introduction}
Optical lattice formed by interference of counter propagating laser
beams is one of the most fruitful technical inventions in the progress
of atomic gas physics~\cite{FGOL}. 
The seminal reports (see, e.g., Ref.~\cite{quantum simulator}) indicate
that this system is regarded as a quantum simulator to 
emulate electronic structures in solid-state systems, with
controllability of model parameters and flexibility of lattice geometric
structures. 
The interaction is widely tuned from
strongly-attractive to repulsive couplings by the Feshbach resonance. 
The different lattice structures, such as a 1D chain, bipartite
square 2D lattice\,\cite{bipartite} like High-$T_{\rm c}$ cuprate
superconductors, and frustrated triangular lattices\,\cite{triangular} 
are available by tuning laser interference.  

\begin{figure}[h]
\begin{center}
\includegraphics[width=1.0\linewidth]{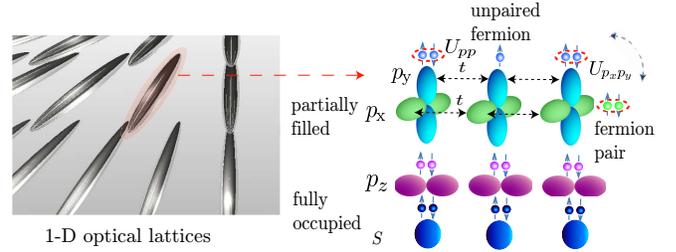}
\end{center}
\caption{(Color Online) Schematic diagram of fermionic gases on an optical
 lattice, with multiple bands (multi-band Hubbard chain). 
The intra-orbital interaction ($U_{pp}$) forms fermion pairs. 
The fermion pairs hop between different orbitals, by the pair-hopping
 interaction ($U_{p_x p_y}$) [see Eq.~(\ref{eq:p1d})].}
\label{schematic fig}
\end{figure}

Simulating solid-state electronic structures in optical lattices
requires treating the orbital degrees of freedom, as well as the charge
and the spin degrees. 
Orbital degeneracy plays a crucial role in transition metals, for
example.  
Such materials are currently targets in applied physics, owing to their
wide usage of various industrial scene.  
Ultracold atomic gases with multiple band-degeneracy enable us to 
directly address quantum phenomena associated with the orbital
degrees~\cite{bipartite,p-band,spin3/2,orbital order,Yb,Tsuchiya;Paramekanti:2012,Zi Cai,Kobayashi,topological ladder}. 
In this paper, focusing on $p$-orbitals next higher to the lowest
$s$-orbital, we show that the low-lying double degenerate orbitals lead
to a rich phase structure of the ground states in an attractively
interacting 1D chain (see Fig.~\ref{schematic fig}).  
All the numerical calculations are done by the density matrix
renormalization method (DMRG) (See, e.g., Refs.~\cite{DMRG1,DMRG2}).  
We also derive an effective model to clarify the origin of the resultant
quantum phase. 

To identify the ground state in a many-body system is a primary issue
for understanding quantum many-body effects. 
The nature of the ground state depends on the degeneracy
intrinsic to a many-body system. 
In our system, the $p$-orbital degeneracy is a key ingredient of the
various quantum phases.  
The $p$-orbitals in a 1D chain along $z$-axis lead to double degeneracy with
respect to $p_{x}$ and $p_{y}$ orbitals, as seen 
in Fig.~\ref{schematic fig}. 
%%%A schematic figure%%%

In this paper, we list up the ground-state properties in this attractive
1D chain. 
First, we show that the \textit{inter-orbital} interaction leads to the
emergence of the Haldane phase~\cite{Haldane}, close to half-filling. 
Below half-filling, the Luther-Emery
phase~\cite{Luther} occurs, in the same way as
\textit{single-band} attractive Hubbard chains~\cite{Machida_attra,Gao}. 
In contrast to the gapless charge excitations in the Luther-Emery phase,
the Haldane phase brings about a charge gapped feature.  
We remark that the Haldane phase with a charge excitation gap, a
nonlocal string order, and edge states is known as a Haldane insulator
phase~\cite{Haldane boson1,Haldane boson2,Haldane boson3,Haldane boson4,
Haldane fermi1,Haldane fermi2}. 
This realization is proposed in bosonic chains with dipole
interaction~\cite{Haldane boson1,Haldane boson2,Haldane boson3,Haldane
boson4} and multi-component fermionic chains~\cite{Haldane
fermi1,Haldane fermi2}. 
A two-leg spin-$1/2$ ladder also shows the presence of a
similar gapped phase~\cite{Hund}. 
Our numerical calculations and effective model reveal the occurrence of
such an intriguing insulator phase in the present system. 
Next, in the presence of population imbalance, we show that a phase
separation of polarized components occurs in a low fermion-pair-density
case. 
This behavior comes from the double exchange interaction~\cite{double
exchange1,double exchange2} between Fermion pairs.  
Finally, we study an effect of trap potential. 
We propose that the trapped system allows the direct verification of the
Haldane gap and the phase-separated polarized phase. 

This paper is organized as follows. 
The $p$-orbitals 1D Hubbard Hamiltonian is derived in
Sec.~\ref{sec:hamiltonian}. 
%In Sec.~\ref{sec:results}A, applying the DMRG method, 
%we solve this Hamiltonian, and we show the presence of the charge gap
%and the edge state. 
In Sec.~\ref{sec:results}A, applying the DMRG method to this
model, we calculate the ground-state of the $p$-orbitals 1D chain, 
and we show the presence of the charge gap
and the edge state. 
Furthermore, deriving an effective spin-$1$ Hamiltonian, 
the realization of the Haldane insulator phase is more evident. 
Section \ref{sec:results}B shows the results with population imbalance. 
We suggest a phase separation of polarized components. 
This phenomenon is explained, in terms of the double exchange
interaction between fermion-pair particles. 
The effect of harmonic trap potential is shown in
Sec.~\ref{sec:results}C. 
Section \ref{sec:summary} is devoted to the summary.

\section{Model}
\label{sec:hamiltonian}
We start with the Hamiltonian for two-component Fermi gases,  
\be
&&H=
\sum_{\sigma=\uparrow,\downarrow}\int d^{3} \bo{x}
\left(
\psi_{\sigma}^{\dagger}h\psi_{\sigma}
+\frac{g}{2}\psi_{\sigma}^{\dagger}\psi_{\bar{\sigma}}^{\dagger}
\psi_{\bar{\sigma}}\psi_{\sigma}
\right),
\ee
with 
\(
h= -(\hbar^{2}/2m)\nabla^{2}+V_{\rm opt}(\bo{x})
\) and two body interaction $g$. 
The optical lattice potential is 
$V_{\rm opt}(\bo{x})=\sum_{\alpha=x,y,z}V_{\alpha}\sin^2(2\pi \alpha/\lambda_\alpha)$\,. 
When the lattice potential is highly elongated along $z$-axis (i.e.,
$V_x=V_y\gg V_z$ and $\lambda_x=\lambda_y\neq\lambda_z$), this 3D atomic
gases can be decomposed into an array of independent 1D chains, as seen
in Fig.1. 
Throughout this paper, we focus on the case when the multiple higher
orbitals are partially filled and the lower orbitals are fully occupied,
inside each piece well of the optical lattices. 
Such a high-density filling is attainable, tuning either the total
particle number or the confinement of the harmonic trap potential. 
Amoung the $p$-orbitals, the Bloch band formed by $p_{z}$ (i.e., a
component along the elongated direction) has a different character from
$p_{x}$ and $p_{y}$, as shown in Fig.~\ref{schematic fig}. 
Hence, we focus on the double degenerate $p_x$- and $p_y$-orbitals,
hereafter. 

Now, we derive a 1D Hamiltonian with $p$-orbital degeneracy. 
We first approximate the optical lattice potential,   
$
V_{\rm opt}(\bo{x})\simeq 
V_{z}\sin^2(2\pi z/\lambda_z)+
\sum_{\alpha=x,y}V_{\alpha}(2\pi \alpha/\lambda_\alpha)^2
$. 
Then, we expand the field operator as 
\begin{equation}
\psi(\bm{x})
=\sum_{i}\sum_{p=p_x,p_y}
c_{p,\sigma,i}\,u_{p}(\bm{x}_{\bot})w_{i}(z),
\end{equation}
with two kinds of the functions $u_{p}$ and $w_{i}$. 
The former is the exact solution of 
$h_{\bot}u_{p}=\epsilon_p u_{p}$, with 
$h_{\bot}=[-(\hbar^2 /2m) \nabla_{\bot}^{2} +\sum_{\alpha=x,y}V_{\alpha}(2\pi
\alpha/\lambda_\alpha)^2]$, and the latter is the Wannier function
formed by the optical lattice potential. 
Using the tight-binding approximation, we obtain
%\begin{widetext}
\be
H=\sum_{p,\sigma}
\sum_{<i,j>}h_{p,\sigma,i,j}^{({\rm t})}+
\sum_{p,\sigma,i}h_{p,\sigma,i}^{(\mu)}+
\sum_{p,p',i}h_{p,p',i}^{({\rm U})}\,, \lb{eq:p1d}
\ee
with 
%\be
%&&
%h_{p,\sigma,i,j}^{({\rm t})}=-t
%c_{p,\sigma,i}^{\dagger}c_{p,\sigma,j}\,, \nn \\
%&&
%h_{p,\sigma,i}^{(\mu)}=-\mu n_{p,\sigma,i}\,, \nn\\
%&&
%h_{p,p',i}^{({\rm U})}
%=U_{pp'}
%\Big{ [ }
%n_{p,\uparrow,i}n_{p',\downarrow,i} 
%+
%\Big{ ( }
%c_{p,\uparrow,i}^{\dagger}c_{p,\downarrow,i}^{\dagger}
%c_{p',\downarrow,i}c_{p',\uparrow,i} \nn\\
%&&\qquad\qquad\qquad
%+c_{p,\uparrow,i}^{\dagger}c_{p',\downarrow,i}^{\dagger}
%c_{p,\downarrow,i}c_{p',\uparrow,i} 
%\Big{ ) }(1-\delta_{pp'})
%\Big{ ] } \nn\,,
%\ee
\be
&&
h_{p,\sigma,i,j}^{({\rm t})}=-t
c_{p,\sigma,i}^{\dagger}c_{p,\sigma,j}\,, \nn \\
&&
h_{p,\sigma,i}^{(\mu)}=-\bar{\mu} n_{p,\sigma,i}\,, \nn\\
&&
h_{p,p'=p,i}^{({\rm U})}
=U_{pp}
\left(n_{p,\uparrow,i}-\frac{1}{2}\right)
\left(n_{p,\downarrow,i}-\frac{1}{2}\right)\,, \nn \\
&&h_{p,p'\neq p,i}^{({\rm U})}
=U_{pp'}
\Big{ ( }
\bo{\rho}_{p,i}\cdot\bo{\rho}_{p',i}
-\bo{S}_{p,i}\cdot\bo{S}_{p',i}
\Big{ ) }\,, \nn 
\ee
where $\bar{\mu}=\mu-(U_{pp}+U_{p_xp_y})/2$. 
The hopping and the on-site interaction energy integrals are defined by,
respectively,  
\begin{eqnarray}
&&
t=-\int dz\,
w_{i+1}
\Big{(}
\frac{-\hbar^2}{2M}\frac{d^2}{d z^2}
+V_{z}\sin^{2} \frac{2\pi z}{\lambda_z} \Big{)}\, w_{i}, \\
&&
U_{pp'}=g\int d^{3}\bo{x}\,w_{i}^{4}u_{p}^2u_{p'}^2 .
\end{eqnarray}
The on-site number density of $p$ orbital is 
\mbox{
\(
n_{p,\sigma,i}=
c_{p,\downarrow,i}^\dagger c_{p,\uparrow,i}
\)}. 
The spin-$1/2$ operator is 
\mbox{
\(
\bo{S}_{p,i}
=
\frac{1}{2}\sum_{\sigma,\sigma'}
c_{p,i,\sigma}^{\dagger}\bo{\tau}_{\sigma,\sigma'}
c_{p,i,\sigma'}
\)}, with the $2\times 2$ Pauli matrices 
$\bo{\tau}=(\tau^{(x)},\,\tau^{(y)},\,\tau^{(z)})$, whereas the pseudo
spin-$1/2$ one is defiend as  
\be
&&\rho_{p,j}^{(x)}=\frac{1}{2}(\rho_{p,j}^{(+)}+\rho_{p,j}^{(-)})\,, \quad
\rho_{p,j}^{(y)}=\frac{1}{2i}(\rho_{p,j}^{(+)}-\rho_{p,j}^{(-)})\,, \nn \\
&&\rho_{p,j}^{(z)}=\frac{1}{2}\Big(\sum_{\sigma}n_{p,\sigma,j}-1\Big)\,,\nn 
\ee
with 
\(
\rho_{p,j}^{(+)}=c_{p,\uparrow,j}^{\dagger}c_{p,\downarrow,j}^{\dagger}
\)
and 
\(
\rho_{j}^{(-)}=[\rho_{j}^{(+)}]^{\dagger}
\). 
%This pseudo spin-$1/2$ operator is used for describing a fermion-pair
%particle in our discussion. 
Throughout this paper, we use the symbol $U_{pp}$ for representing the
intra-orbital interaction strength, since 
$U_{p_{x}p_{x}} = U_{p_{y} p_{y}}$. 
We also find that $U_{p_x p_y}=U_{p_y p_x}$. 
The intra- and the inter-orbital interaction strength are evaluated by the
exact solution of $h_{\bot}u_{p}=\epsilon_p u_{p}$.  
We can find the relation $U_{p_xp_y}=(4/9)U_{pp}$. 
This relation is also used throughout this paper. 
In this paper, we set $U_{pp^{\prime}}$ as a negative value, i.e., an
attractive two-body interaction $g<0$. 

The virtue of using the spin representation by $\bo{S}_{p,j}$ and
$\bo{\rho}_{p,j}$ in the Hamiltonian is to clarify an underlying
symmetry feature of this model. 
Furthermore, our numerical results are intuitively understood by this
representation; in the subsequent sections a fermion-pair particle will
be discussed, in terms of $\bo{\rho}_{p,j}$. 
Let us here show the symmetry of Eq.~(\ref{eq:p1d}) explicitly. 
Taking the summation over $p_{x}$ and $p_{y}$, we build two kinds of
operators, 
\( 
S_{i}^{(l)}=\sum_{p}S_{p,i}^{(l)}
\)
and 
\(
\rho_{i}^{(l)}=\sum_{p}\rho_{p,i}^{(l)}
\), with $l=x,\,y,\,z$. 
We can obtain the operators for $l = \pm $, in a similar manner to the
definition of $\bo{\rho}_{p,j}$. 
The former is a (local) spin-$1$ operator, whereas the latter is a
(local) pseudo spin-$1$ operator. 
After the straightforward calculations, we obtain the algebraic
relations, 
\be
[H,S^{(l)}]=0\,,\quad
[H,\rho^{(\pm)}]=\mp 2\bar{\mu}\rho^{(\pm)}\,,\quad
[H,\rho^{(z)}]=0\,,
\ee
with $S^{(l)} = \sum_{i}S_{i}^{(l)}$ 
and $\rho^{(l)} = \sum_{i}\rho_{i}^{(l)}$. 
The first one indicates that Hamiltonian (\ref{eq:p1d}) is
isotropic with respect to the global spin rotation. 
The latter two relations mean that this model possesses a highly
symmetric property at half filling ($\bar{\mu}=0$). 
In other words, the present Hamiltonian has 
\(
SU(2)_{\rm spin} \times SU(2)_{\text{pseudo-spin}} 
\simeq SO(4)
\) symmetry at $\bar{\mu}=0$. 

\section{Results}
\label{sec:results}
Let us study the quantum phases of Eq.~(\ref{eq:p1d}) 
at zero temperature. 
All the numerical calculations are performed by the DMRG
method~\cite{DMRG1,DMRG2}. 
Our DMRG code is directly extended toward ladder systems, by
parallelizing the superblock matrix diagonalization~\cite{YAMADA}. 
The number of states kept is varied from $m=400$ to maximally $1000$
depending on the convergence tendency of the calculations.  
The boundary condition is open in all the calculations. 
The subsequent subsection shows the results in a spatially uniform case,
without any population imbalance. 
Next, we turn into the case with spatially uniformity and population
imbalance. 
In the third subsection, the effect of the confinement harmonic trap
potential is studied. 

\subsection{Zero population imbalance}
\begin{figure}[h]
\begin{center}
\includegraphics[width=1.00\linewidth]{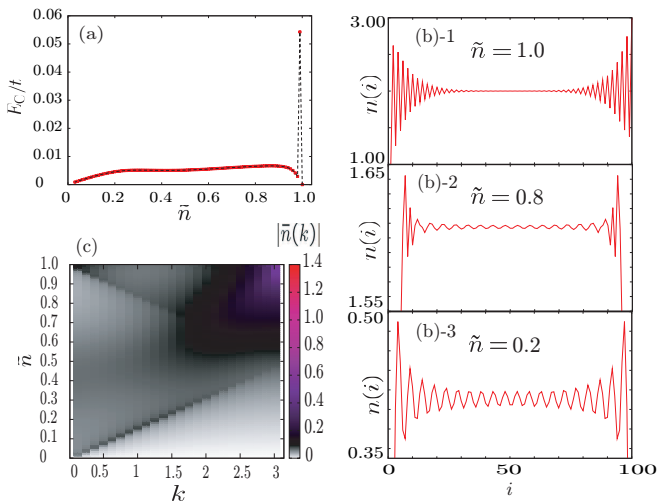}
\end{center}
\caption{(Color online) (a) Charge gap $E_{\rm c}$ versus filling rate
 $\tilde{n}$. 
(b) Particle density $n(i)$, with $\tilde{n}=1.0,\,0.8,\,0.2$.  
(c) Absolute value of the Fourier-transformed density fluctuations, 
 $|\bar{n}(k)|$ on \mbox{$k$--$\tilde{n}$} plane.  
In all the figures, the population imbalance is $P=0$, the coupling
 parameters are $U_{pp}=-10$ and $U_{p_x p_y}=(4/9)U_{pp}$, and the
 total lattice site number is $L=100$. }
\label{phase1}
\end{figure}
Figure \ref{phase1}(a) shows our DMRG results about the charge gap with
respect to the filling rate $\tilde{n}= (1/2L)\sum_{i}n(i)$, where
$n(i) = \sum_{p,\sigma}n_{p,\sigma,i}$ is the particle density and $L$
is the total lattice site number. 
The charge gap is evaluated by 
\(
E_{\rm C} = E(N+\uparrow\downarrow)+E(N-\uparrow\downarrow)-2E(N)
\), with the DMRG ground-state energy $E(\cdot)$. 
In Fig.~\ref{phase1}, the population imbalance $P$ is zero; 
\(
P \equiv \sum_{p,i}(n_{p,\uparrow,i}-n_{p,\downarrow,i}) = 0
\). 
We find that the charge gap drastically grows up when $\tilde{n}$ is
close to half filling (i.e., $\tilde{n}\simeq 1.0$). 
In contrast, when $\tilde{n}$ decreases from $1$, the gap reduces. 
These behaviors indicate that in this double-degenerate-band attractive
1D model a gapped phase emerges, close to half-filling. 
We stress that the charge excitation gap does not open in single-band
attractive 1D Hubbard chains~\cite{Machida_attra,Gao}.

We show that the emergence of this charge gap is attributed to the
Haldane gap, by mapping the present Hamiltonian onto an interacting
spin-$1$ chain. 
Using the second order perturbation~\cite{BW} and the
attractive-repulsive transformation~\cite{ARtrans}, we obtain an
effective model of Eq.~(\ref{eq:p1d}), for strong coupling regime 
\mbox{$|U_{pp'}|\gg t$}.  
The attractive-repulsive transformation~\cite{ARtrans} makes
Eq.~(\ref{eq:p1d}) a half-filled system. 
This transformation is defined by 
$
c_{p,\uparrow,i}
=
\bar{c}_{p,\uparrow,i}
$ 
and 
$
c_{p,\downarrow,i}=(-1)^i
\bar{c}_{p,\downarrow,i}^\dagger
$. 
For the \textit{free} Hamiltonian $\sum_{p,p',i}h_{p,p',i}^{({\rm U})}$,
we obtain
\be
&&H_{\rm eff}
=
\sum_{p,\sigma,i}
\bar{\mathcal{P}}h_{p,\sigma,i}^{({\rm \mu})}\bar{\mathcal{P}} 
-
\!\!\!\!\!\!\!\!\!
\sum_{p,\sigma,<i,j>\atop  p',\sigma',<i',j'>}
\!\!\!\!\!\!\!\!\!
V_{p,\sigma,i,j} \, H_{0}^{-1} \,
V_{p^{\prime},\sigma^{\prime},i^{\prime},j^{\prime}}^{\dagger},
%
%(\bar{\mathcal{P}}h_{p,\sigma,i,j}^{({\rm t})}\bar{\mathcal{Q}})
%H_{0}^{-1}
%(\bar{\mathcal{Q}}h_{p',\sigma',i',j'}^{({\rm t})}\bar{\mathcal{P}})\,, \nn\\
\lb{eq:BW}
\ee
with 
\(
V_{p,\sigma,i,j}
=
\bar{\mathcal{P}}h_{p,\sigma,i,j}^{({\rm t})}\bar{\mathcal{Q}}
\). 
The operators $\bar{\mathcal{P}}$ and $\bar{\mathcal{Q}}$ are,
respectively, the projectors onto the subspaces, 
\be
\mathcal{\bar{H}}_{P}&=&\otimes_{i}
\big{\{}
|\bar{\uparrow},\bar{\uparrow}\rangle\,,
|\bar{\downarrow},\bar{\downarrow}\rangle\,,
(|\bar{\downarrow},\bar{\uparrow}\rangle+
|\bar{\uparrow},\bar{\downarrow}\rangle)/\sqrt{2}
\big{\}} \\
\mathcal{\bar{H}}_{Q}&=&\otimes_{i}
\big{\{}
|\bar{\uparrow},\bar{0}\rangle\,, 
|\bar{0},\bar{\uparrow}\rangle\,, 
|\bar{\downarrow},\bar{0}\rangle\,, 
|\bar{0},\bar{\downarrow}\rangle\,, \nn \\
&&\quad|\bar{\uparrow},\bar{\uparrow}\bar{\downarrow}\rangle\,, 
|\bar{\uparrow}\bar{\downarrow},\bar{\uparrow}\rangle\,, 
|\bar{\downarrow},\bar{\uparrow}\bar{\downarrow}\rangle\,, 
|\bar{\uparrow}\bar{\downarrow},\bar{\downarrow}\rangle
\big{\}}\,.
\ee 
Here, $|\bar{\cdot},\bar{\cdot}\rangle$ means 
$|\bar{\cdot},\bar{\cdot}\rangle = 
|\bar{\cdot}\rangle_{p_{x}} |\bar{\cdot}\rangle_{p_{y}}$ and 
the ket vector $|\bar{0}\rangle_{p_{x(y)}}$ is defined by
$\bar{c}_{p_{x(y)},\sigma,i}|\bar{0}\rangle_{p_{x(y)}}=0$. 
Eq.~(\ref{eq:BW}) can be rewritten in terms of the pseudo-spin-$1$
operators 
$\rho_{i}^{(\pm)}$ and
$\rho_{i}^{(z)}$\,.
%These operators represent a fermion-pair particle. 
The effective low-energy Hamiltonian 
of the system is reduced to  
a 1D pseudo-spin-$1$ chain, 
\be
H_{\rm eff}&=&
J_{\rm ex}\sum_{<i,j>}\Big{ [}
\rho_{i}^{(z)}\rho_{j}^{(z)}-
\frac{1}{2}(\rho_{i}^{(+)}\rho_{j}^{(-)}+\rho_{i}^{(-)}\rho_{j}^{(+)})
\Big{ ]}\nn \\
&&-\sum_{i}2\bar{\mu}\rho_{i}^{z}\,,\lb{eq:xxz}
\ee 
with $J_{\rm ex}=2t^2/\left(|U_{pp}|+|U_{p_x p_y}|\right)$\,.  
Thus, we find that the charge gap (i.e., the pseudo spin gap) opens,
according to Haldane's conjecture\,\cite{Haldane}. 
 
Next, we show another evidence of the Haldane phase in our system. 
In the open boundary condition, it is well-known that the Haldane phase
forms a free half spin state near the boundaries.  
Figure \ref{phase1}(b)-1 shows that a staggered charge density
modulation occurs and exponentially decays toward the bulk region. 
This behavior corresponds to the $S=1/2$ edge
state~\cite{AKLT,Miyashita}. 
Thus, a free half spin emerges as a free fermion-pair particle, near
the edges. 
This free fermion-pair particle induces a \textit{gapless} charge
excitation when the system is the half filling case. 
Figure \ref{phase1}(a) shows this behavior; the charge gap occurs right
below half filling, whereas a gapless behavior is found at
$\tilde{n}=1.0$. 
We stress that the gapless behavior at $\tilde{n}=1.0$ comes from the
edge contribution. 
%We can find that in the bulk region the charge gap still opens, like the
%case at $\tilde{n}=0.99$. 

%%%%%%%%%%% 1/2  %%%%%%%%%%%%%%%%%
Now, we study the case when the filling rate is much lower than half
filling.  
Figures \ref{phase1}(b)-2 and (b)-3 show that the periodic
oscillations dominate over the whole spatial region, not only the
boundaries, when going below the half filling rate. 
We obtain the charge density wave (CDW) below half filling. 
The spatial periodicity indicates the presence of the Luther-Emery phase. 
If the Luther-Emery phase occurs, ths periodicity of the CDW should be
characterized by the Fermi wave vector in the equivalent spinless 
Fermion~\cite{LL1,1Dbook}, 
$2k_{\rm F}=2\pi(\rho-|\sum_{i}\rho_i^{(z)}|/L)$. 
Here,  $\rho$ is pseudo-spin length (i.e., $\rho=1$). 
Let us calculate the Fourier transformed density fluctuations, 
\(
\bar{n}(k)= \sum_j [n(i)-2\tilde{n}] e^{ikj}/\sqrt{L}
\). 
Figure \ref{phase1}(c) shows the absolute value  
\(
|\bar{n}(k)|
\), varying $k$ and $\tilde{n}$. 
We find the two kinds of the peaks, a strong peak around
$(k,\tilde{n})\simeq (3.14,1.0)$ caused by staggered CDW (edge states)
and  a peak consistent with the prediction of the Luttinger theory
$k=2k_{\rm F}=2\pi\tilde{n}$\,. 
The latter peak indicates the emergence of the Luther-Emery phase below
half-filling. 
In other words, our model below half filling behaves like an attractive 1D
Hubbard chain. 

\subsection{Nonzero population imbalance}
\begin{figure}[htbp]
\begin{center}
\includegraphics[width=1.00\linewidth]{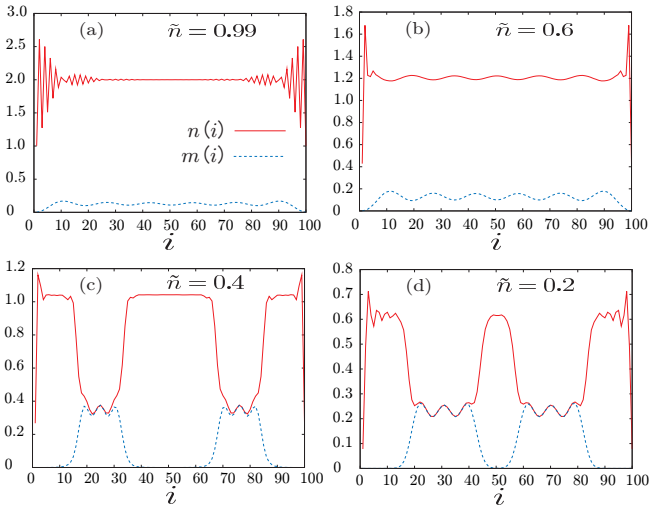}
\end{center}
\caption{(Color online) Spatial distributions of the particle density
 $n(i)$ (soild line) and the spin density $m(i)$ (dashed line), with
 filling rates (a) 
 $\tilde{n}=0.99$, (b) $\tilde{n}=0.6$, (c) $\tilde{n}=0.4$, and (d)
 $\tilde{n}=0.2$. 
In the all figures, the population imbalance is fixed as
 $P=12$. 
Other physical parameters are the same as in Fig.\ref{phase1}. 
}
\label{open2}
\end{figure}
We study the effects of population imbalance in a spatially uniform
case. 
Figure \ref{open2} shows the spatial distributions of the particle
density $n(i)$ and the spin density 
\(
m(i) = \sum_{p}(n_{\uparrow,p,i}-n_{\downarrow,p,i})
\), with fixed population imbalance $P=12$, varying filling rates. 
At higher filling rates ($\tilde{n}=0.99,\,0.6$), we obtain the spin
density wave, as seen in Figs.~\ref{open2}(a,b) (dashed lines).  
We find that the spatial period is characterized by 
\(
2\Delta k_{{\rm F},p} = \pi P/L 
\) (i.e., 
\(
2\Delta k_{{\rm F},p}L/ 2\pi
= 6 
\)), with the difference of the $p$-orbital Fermi wave vectors 
\(
\Delta k_{{\rm F},p} 
=
(\pi/L)\sum_{i}(
n_{p,\uparrow,i} - n_{p,\downarrow,i}
)
\). 
%As for $\tilde{n}=0.99$ and $\tilde{n}=0.6$, the spin density wave 
%with the periodicity $2\Delta k_{{\rm F},p}=\pi P/L$ is found in the
%spin distribution profiles, where 
%$\Delta k_{{\rm F},p}$ is the difference of the $p$-orbital Fermi wave
%vectors 
%$\Delta k_{{\rm
%F},p}=(\pi/L)\sum_{i}(n_{p,\uparrow,i}-n_{p,\downarrow,i})$. 
The density profile depends on the filling rate more sensitively. 
Right below half filling ($\tilde{n}=0.99$), a uniform density profile
is found in the bulk region, and a staggered CDW occurs at the
edges. 
These results are similar to the case without population imbalance; the
charge gap opens in the bulk region, while a free half spin state
may induce gapless charge excitations near the boundaries. 
When the filling rate decreases slightly ($\tilde{n}=0.6$), a small
spatial modulation is found in the whole spatial region. 
%When $\tilde{n}=0.6$, we find that a $2\Delta k_{{\rm F},p}$ oscilaltion
%occurs in the whole spatial region. 
At much lower filling rates ($\tilde{n}=0.4,\,0.2$), we find a drastic
effect of the population imbalance. 
%The most drastic effect of the population imbalance cases is found at
%$\tilde{n}=0.4$ and $\tilde{n}=0.2$\,. 
Figures \ref{open2}(c,d) show a phase separation of polarized
components. 
The small spatial modulation of the particle density for $\tilde{n}=0.6$
[Fig.~\ref{open2}(b)] is regarded as a precursory phenomenon of this
phase separation. 
Decreasing the filling rate enhances the amplitude of the
density-profile oscillation. 
Then, the low-particle-density regions are created for lower filling
rates~\cite{note1}. 
%This phase separated behavior is prominent for the strong attractive interaction $U_{pp}/t$\,. 
%Fig.\ref{2exchange}.(a) show the result via the change of $U_{pp}$ and $\tilde{n}$ 
%with the fixed $P=12$ and $L=100$\,. 

\begin{figure}[h]
\begin{center}
\includegraphics[width=1.00\linewidth]{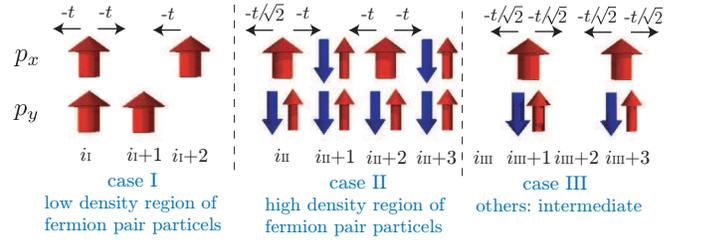}
\end{center}
\caption{(Color online)  Schematic diagram of hopping processes of
 unpaired fermions (single up-arrow) in fermion-pair particles (pair of
 up- and down-arrow). }
\label{2exchange}
\end{figure}

To clarify the origin of the phase separation, we focus on the kinetics
energy of unpaired fermions. 
Applying the first order perturbation to Eq.~(\ref{eq:p1d}), 
we obtain
\be
%H_{\rm K}=-t\sum_{p,\sigma}\sum_{<i,j>}
%\mathcal{P}_i c_{p,\sigma,i}^{\dagger}\mathcal{P}_i
%\mathcal{P}_j c_{p,\sigma,j}\mathcal{P}_j\,,
H_{\rm K}=-t\sum_{p,\sigma}\sum_{<i,j>}
\mathcal{P}_i\mathcal{P}_j 
h_{p,\sigma,i.j}^{({\rm t})}
\mathcal{P}_i\mathcal{P}_j .
\ee 
The operator $\mathcal{P}_i$ is the projector onto the
subspace spanned by the nonzero spin imbalance states and the
pseudo-spin-$1$ states at the spatial site $i$, 
\be
\mathcal{H}^{(i)}_{P>0,\,\rho=1} 
&=& 
\Big{\{}
|\uparrow,\uparrow\downarrow\rangle\,, 
|\uparrow\downarrow,\uparrow\rangle\,, 
|\uparrow,0\rangle\,, 
|0,\uparrow\rangle \,, 
|\uparrow,\uparrow\rangle , \nn \\
&&
\quad
|0,0\rangle\,,
|\Psi_{+}\rangle\,,
|\uparrow\downarrow,\uparrow\downarrow\rangle,
\Big{\}}\,,
\ee 
with 
\mbox{
\(
|\Psi_{+}\rangle 
=
(|\uparrow\downarrow,0\rangle+
|0,\uparrow\downarrow\rangle)/\sqrt{2}
\)}. 
The ket vector symbol $|\cdot,\cdot\rangle$ means 
$|\cdot,\cdot\rangle =  
|\cdot\rangle_{p_{x}} |\cdot\rangle_{p_{y}}$, with 
\mbox{$c_{p_{x(y)},\sigma,i}|0\rangle_{p_{x(y)}}=0$}.  
Figure \ref{2exchange} shows hopping processes of the unpaired Fermion
(single up-arrow) embedded in the continuum formed by the fermion-pair particles
(pair of up- and down-arrows). 
If a low fermion-pair-particle-density region spreads, as seen in case I
of Fig.~\ref{2exchange}(b), the transfer probability of the unpaired
Fermions is $-t$. 
This fact is confirmed by, for example, 
\mbox{
\(
\,_{i_{\rm I}+1}\langle \uparrow,0 |\,_{i_{\rm I}}\langle 0,\uparrow| 
H_{\rm K}
|\uparrow,\uparrow\rangle_{i_{\rm I}}|0,0\rangle_{i_{\rm I}+1}
=
-t
\)}. 
Similarly, in a high fermion-pair-particle-density region [see case
\mbox{II} in Fig.~\ref{2exchange}(b)], the transfer probability is
$-t$, since, for example, 
\mbox{
\(
\,_{i_{\rm II}+1}\langle  \uparrow,\uparrow\downarrow |
\,_{i_{\rm II}}\langle \uparrow\downarrow,\uparrow\downarrow| 
 H_{\rm K}
|\uparrow,\uparrow\downarrow\rangle_{i_{\rm II}}
|\uparrow\downarrow,\uparrow\downarrow\rangle_{i_{\rm II}+1}
=-t
\)}. 
In contrast, when the fermion-particle density is intermediate
[e.g. CDW-like configuration as case \mbox{III} of
Fig.~\ref{2exchange}(b)], 
the transfer probability changes, since, for example, 
\mbox{
\(
\,_{i_{\rm III}+1} \langle  \Psi_{+}   |
\,_{i_{\rm III}}   \langle  \uparrow,0 |
H_{\rm K}
|0,0\rangle_{i_{\rm III}}
|\uparrow,\uparrow\downarrow\rangle_{i_{\rm III}+1}
=
-t/\sqrt{2}
\). 
}
Thus, the unpaired Fermions prefer to either low or high
fermion-pair-particle-density regions. 

Now, we apply the above arguments to our numerical results. 
The second order perturbation terms in Eq.~(\ref{eq:xxz}) vanish
as $J_{\rm ex} \to 0$. 
Furthermore, since the lower filling ($\tilde{n}\ll1$) means increasing
the pseudo magnetic field $2\bar{\mu}$, the pseudo spin-spin interaction
of Eq.~(\ref{eq:xxz}) is irrelevant to the total energy. 
Thus, the kinetic energy of the unpaired Fermions is predominant, for strong
attractive interaction $|U_{pp}|\gg t$ and a lower filing rate
($\tilde{n} \ll 1$). 
The results shown in Fig.~\ref{open2} corresponds to the case below half
filling (i.e., a low filling case).  
Therefore, from the consideration about the kinetic energy, the unpaired
Fermions in this figure prefer to the low fermion-pair-particle-density
regions. 
In other words, the spin imbalance excludes the fermion-pair particles 
and leads to a local spin polarized state. 
This process can be regarded as double exchange
interaction~\cite{double exchange1,double exchange2} between the
fermion-pair particles. 

\subsection{Harmonic trap potential}
\begin{figure}[h]
\begin{center}
\includegraphics[width=1.00\linewidth]{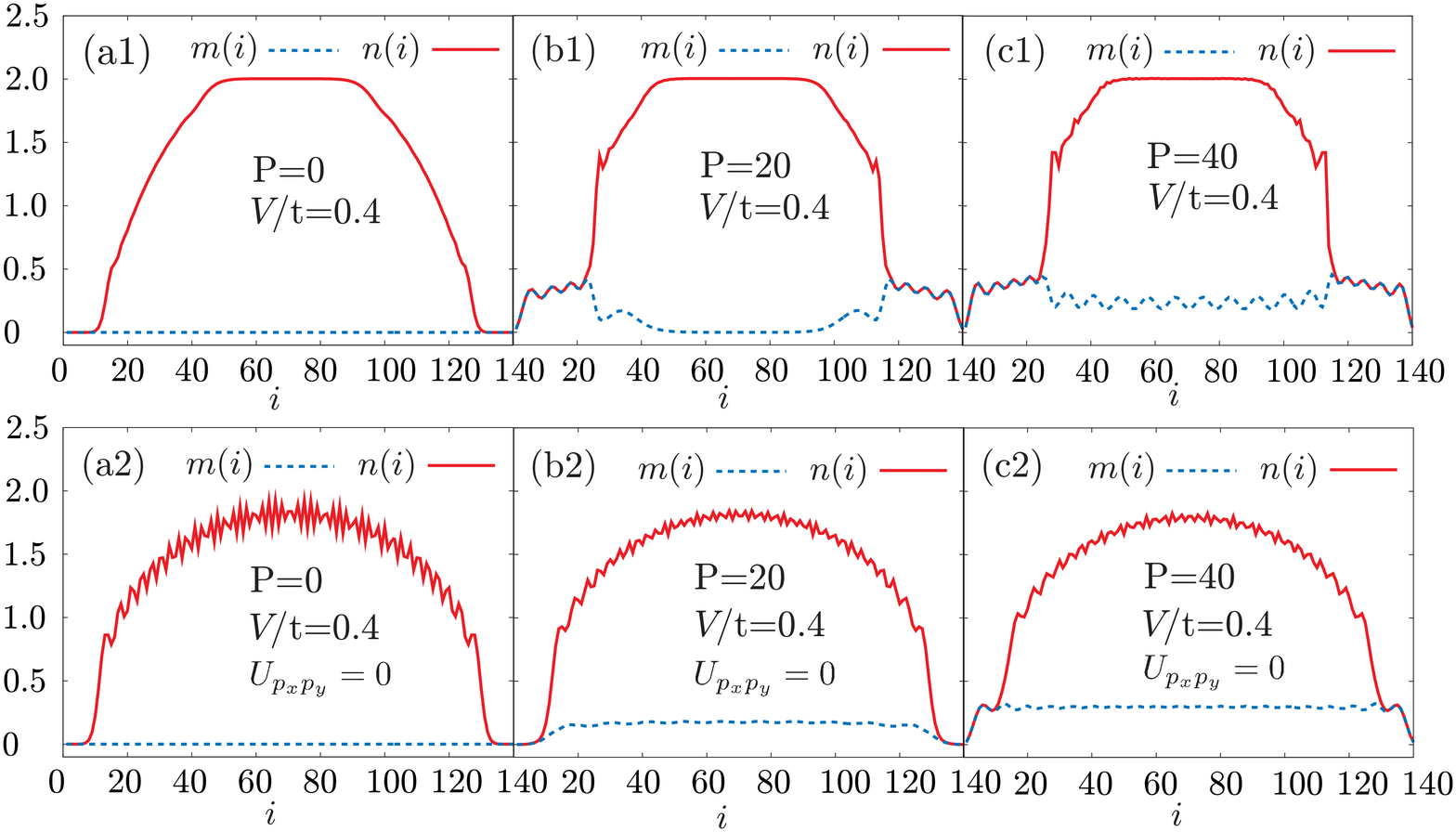}
\end{center}
\caption{(Color Online) Spatial distributions of the particle density
 $n(i)$ (solid line) and the spin density $m(i)$ (dashed line) in
 harmonic trap potential, with different population imbalances $P=0$, 
 $20$, $40$. 
The trap potential strength is $V/t=0.4$. 
The total particle number is $180$. 
The upper panels (a1)--(c1) show the results for the present double
 degenerate $p$-orbital 1D chain.  
For comparison, we show in the lower panels (a2)-(c2) the results for
 zero intra-orbital interaction ($U_{p_xp_y}=0$), i.e., a single-band
 attractive Hubbard chain. 
}
\label{trap_V=0.4}
\end{figure}
We take the effect of the harmonic trap potential into account.  
The trap potential is typically employed in atomic gas experiments, to
avoid the escape of the atoms. 
When the harmonic trap is considered, we simply add a potential term to our
model. 
Thus, the total Hamiltonian is 
\mbox{$H+\sum_{p,\sigma,i}V_{\rm ho}(i)n_{p,\sigma,i}$}, 
with \mbox{$V_{\rm ho}(i)=V[2/(L-1)]^2 [i-(L+1)/2]^2$}. 
We obtain the ground state of this modified Hamiltonian, using the DMRG
method. 

%First, we focus on the case of $V/t=0.2$\,, 
%where particle density do not exceed $n(i)=2$(half filling)\,. 
%Figure \ref{trap_V=0.4}(a1) shows the results, without population
%imbalance ($P=0$). 
%The significant feature of the trapped system is the emergence of the
%Mott core. 
%This result is a contrast to the case of single-band attractive Fermi
%gases, as seen in Fig.~\ref{trap_V=0.4}(a2); there is no Mott-core
%structure. 
%This structure is clear evidence of the formation of the Haldane
%insulator phase at half filling $n(i)=2$.  
%The edge state of the Haldane phase is not observed in our trapped
%system. 
%The effect of the trap potential is considered to be a spatially
%dependent pseudo magnetic field, $\bar{\mu}(i)=\bar{\mu}+V(i)$, in terms
%of effective Hamiltonian (\ref{eq:xxz}). 
%The smooth change of this magnetic field leads to the disappearance of
%the staggered CDW. 
%Below half filling ($n(i)=2$), the present system is expected to show
%the CDW oscillation as discussed in section III.A. 
%However, the density oscillation is not sharply observed in
%Fig.~\ref{trap_V=0.4}.(a1). 
%The 1D single-band case shows a clear CDW oscillation
%[Fig.~\ref{trap_V=0.4}(a2)]. 
%Therefore, the CDW order in the double degenerate $p$-orbital system is
%weaker, compared to the single band case. 
Figure \ref{trap_V=0.4}(a1) shows the results, without population
imbalance ($P=0$). 
The significant feature of the trapped system is the emergence of the
Mott core. 
This result is a contrast to the case of single-band attractive Fermi
gases, as seen in Fig.~\ref{trap_V=0.4}(a2); there is no Mott-core
structure. 
This structure implies that the Haldane insulator phase is formed at
half filling $n(i)=2$. 
From the viewpoint of Eq.~(\ref{eq:xxz}), the effect of the trap
potential is regarded as spatially-dependent pseudo-magnetic field,
$\bar{\mu}(i)=\bar{\mu}-V(i)$. 
Thus, the Mott-core is considered to be a ``magnetization''
plateau associated with opening of the Haldane gap. 
The edge state of the Haldane phase is not observed in our trapped
system. The smooth change of this magnetic field leads to the disappearance of
the staggered CDW. 
Below half filling ($n(i)=2$), the present system is expected to show
the CDW oscillation as discussed in section III.A. 
However, the density oscillation is not sharply observed in
Fig.~\ref{trap_V=0.4}.(a1). 
The 1D single-band case shows a clear CDW oscillation
[Fig.~\ref{trap_V=0.4}(a2)]. 
Therefore, the CDW order in the double degenerate $p$-orbital system is
weaker, compared to the single band case. 

Next, we study the results in the presence of population imbalance. 
Figures \ref{trap_V=0.4}(b1, c1) show that the Haldane insulator phase
occurs, in the same way as the zero population imbalance. 
Similarly, we find that there is no Mott-core structure, therefore no
Haldane insulator phase occurs in a single-band attractive Hubbard chain
[Figs.~\ref{trap_V=0.4}(b2,c2)]. 
The effect of the population imbalance appears, as a phase separation of
the polarized components on merge of the trap potential
[Fig.~\ref{trap_V=0.4}(b1,c1)].  
As discussed in Sec.~\ref{sec:results}\,B, the kinetic energy of the
unpaired Fermions prefers to a low fermion-pair-particle-density region. 
As a result, the polarized components concentrate at the edges of the
trap potential (low density region). 
The phase separation of the polarized components can be 
observed in single band attractive Fermi gases with trap
potential~\cite{Feiguin,Tezuka,phase separation1,phase separation2}
[see Fig.~\ref{trap_V=0.4}(c2), as well]. 
However, the phase separation in a single-band case is vague, compared
to the double degenerate $p$-orbital system, as seen in
Figs.~\ref{trap_V=0.4}(b1,b2). 
The phase separation of the $p$-orbital Fermi gases is strong and 
easily induced by the double exchange interaction between the
fermion-pair particles. 

Summarizing the results for the trapped system, we can suggest the direct
and concrete check of our predictions in experiments. 
%The $p$-band Mott core is the definite evidence of the Haldane insulator
%phase, and is detectable via the measurement of the particle density
%profile. 
The Mott-core structure is induced by the Haldane
gap, and is detectable via the measurement of the particle density
profile. 
The phase separation of the polarized states is identified by comparing
the spin density at the trap center to the one at the trap edges. 
One drawback of the harmonic trap is to erase the edge states related to
the Haldane edge states. 
The realization of box trap potential\,\cite{box trap} may capture such
states. 
Observing the string order parameter gives a strong
signature of the Haldane insulator phase, as well as measuring the
$p$-band Mott core and the edge states. 
We suggest that a single-site addressing technique (see, e.g.,
Ref.~\cite{string order}) allow a direct check of this quantity. 
Thus, calculating the string order parameter on the $p$-band Mott-core
is an interesting future work. 

\section{Summary}
\label{sec:summary}
We explored the quantum phases in a 1D $p$-orbital Fermi gas with
attractive interaction, via the DMRG calculations and the mapping onto
an effective spin-$1$ model. 
To tune the filling rate and the population imbalance induces different
phases, including the Haldane insulator phase, the Luther-Emery 
phase, and the phase separation of the polarized components. 
We also examined the effect of the harmonic trap potential. 
We found the emergence of the Mott core structure (i.e., Haldane
insulator phase), in spite of attractive Fermi gases. 
Moreover, the strong phase separation induced by the population imbalance 
appears at the edges of the trap potential. 
Thus, the trapped system allows the direct verification of our
predictions. 

\begin{acknowledgments}
We wish to thank Y.~Nagai for useful discussions. 
This research was partially supported by a Grant-in-Aid for Scientific Research from JSPS (Grant No. 23500056). 
This work was partially supported by the Strategic Programs for Innovative Research, MEXT, and the Computational Materials Science Initiative (CMSI), Japan. We are indebted to T. Toyama for his support. 
The numerical work was partially performed on Fujitsu BX900 in JAEA. We acknowledge support from the CCSE staff.
\end{acknowledgments}

\appendix

%\newpage %Just because of unusual number of tables stacked at end


\begin{thebibliography}{99}
\bibitem{note1}
Although the phase separation of polarized components occurs 
in a uniform system at a low filling rate, the spatial configuration is
	not fixed. 
Indeed, we can find that the place of a low-particle-density region
	depends on the system size $L$. 
In the case of a trapped system, the inhomogeneity of the harmonic trap
	potential determines the spatial configuration of the phase
	separations.
\end{thebibliography}
\end{document}